\documentclass[runningheads]{llncs}

\usepackage{xcolor}
\usepackage{amsmath}
\usepackage{amsfonts}
\usepackage{mathtools}
\usepackage{hyperref}
\usepackage[binary-units]{siunitx}
\usepackage{wrapfig}
\usepackage{multirow}

\usepackage[ruled, vlined, linesnumbered]{algorithm2e}

\def\Rset{\ensuremath{\mathbb R}} 
\def\R{\ensuremath{\mathcal R}} 
\def\P{\ensuremath{\mathcal P}} 

\definecolor{jd_blue0}{rgb}{0.51,0.34,1.00}
\definecolor{jdbishop}{rgb}{0.27,0.14,0.29}
\definecolor{jddarkbl}{rgb}{0.18,0.14,0.29}
\definecolor{jd_brown}{rgb}{0.29,0.14,0.14}
\definecolor{jd_green}{HTML}{096F35}
\definecolor{jdorange}{rgb}{1.00,0.55,0.00}
\definecolor{jdredred}{rgb}{0.50,0.00,0.00}
\definecolor{SchoolColor}{rgb}{0.145,0.666,1}
\definecolor{pheniics}{HTML}{1a899c}
\definecolor{pheniics_purple}{HTML}{64003d}
\definecolor{chaptercolor}{gray}{0.8}
\definecolor{bordeau}{rgb}{0.3515625,0,0.234375}

\title{Making use of supercomputers in financial machine learning}

\author{Philippe Cotte\inst{1} \and Pierre Lagier\inst{2} \and
Vincent Margot\inst{1} \and Christophe Geissler\inst{1}}

\authorrunning{P. Cotte et al.}

\institute{
Advestis, Paris, France,
\email{\{pcotte, vmargot, cgeissler\}@advestis.com}, \url{https://www.advestis.com}\\\and
Fujitsu Systems Europe, France,
\email{pierre.lagier@fujitsu.com}}

\begin{document}

\maketitle
\begin{abstract}
This article is the result of a collaboration between Fujitsu and Advestis.
This collaboration aims at refactoring and running an algorithm based on systematic exploration producing investment recommendations on a high-performance computer of the Fugaku type \cite{fugaku}, to see whether a very high number of cores could allow for a deeper exploration of the data compared to a cloud machine, hopefully resulting in better predictions.


We found that an increase in the number of explored rules results in a net increase in the predictive performance of the final ruleset. Also, in the particular case of this study, we found that using more than around 40 cores does not bring a significant computation time gain. However, the origin of this limitation is explained by a threshold-based search heuristic used to prune the search space. We have evidence that for similar data sets with less restrictive thresholds, the number of cores actually used could very well be much higher, allowing parallelization to have a much greater effect.

\end{abstract}

\section{Introduction}
At a time of rapid development of deep networks, driven by constantly expanding computational capabilities, it is easy to forget a technique that was at the forefront of artificial intelligence in the 1980s: expert systems. Expert systems were based on logical inference capabilities operating on business rules. They were able, by means of dedicated languages such as Prolog, to make automatic deductions from large bases of assertions made by experts, hence the name given to the method.

Over the past decade, deep learning has become established in fields as diverse as molecular synthesis, image recognition and financial system forecasting. The founding idea of expert systems, namely reasoning based on large bodies of human knowledge, has generally been abandoned. However, algorithmic research has continued in the direction of rule-based prediction systems, which differ from expert systems in that the inference rules are obtained by algorithms and not by human experts.

Rule-based prediction systems continue to play an important role in economics and finance. They have at least two interesting properties for practitioners:
\begin{enumerate}
    \item Under conditions of limited complexity, they directly provide interpretable predictions, similar to methods derived from decision trees (such that \emph{CART} \cite{CART}, \emph{ID3} \cite{Quinlan86},
    \emph{C4.5} \cite{Quinlan93} and \emph{RIPPER} \cite{Cohen95}).
    \item Some rule search algorithms do not require the entire data set to be loaded simultaneously into memory. This can be an advantage when resources are limited.
\end{enumerate}

However, despite the apparent simplicity of the output, rule generation algorithms are complex and can lead to intractable computation times if the algorithms are not carefully optimized.

A rule-based algorithm specifically designed for a financial investing process has been presented in \cite{margot2021esg}. This algorithm represented the first version of a generic rule search algorithm developed by Advestis under the name AdLearn.

Due to the combinatorial nature of the search, the presented algorithm was computationally very intensive in its non-optimized version. It did not allow for frequent renewal of the set of rules, a serious drawback given the rapid evolution of the underlying financial phenomena.

Another bottleneck was a high memory consumption during the design of the rules, which limited the number of stocks and features that could be considered in the training. 

These limitations led us to push further the degree of parallelization of the code. In this way, difficulties arise when using complexity reduction heuristics. For example, if the stopping criterion of the algorithm is based on the achievement of a sufficient coverage of the explored space, the result is non-deterministic, being dependent on the physical architecture chosen for the calculation. And clearly, the deterministic nature of the result cannot be negotiated, specially in finance, unless we completely renounce interpretability.

The output of this collaboration between Advestis and Fujistu is a new version of AdLearn designed to run on a massively parallel architecture, which we present in this article.

\section{AdLearn under the lens}
        
We aim at predicting a random variable $Y \in \Rset$, the target, given a random variable $X \in \Rset^d$, the features. To do so, we use an algorithm, named AdLearn, which extracts a \textbf{set of rules} from a learning sample $D_n$.
    
    \subsection{Input Data}

     The learning sample, $(X_i, Y_i)_{1 \leq i \leq n}$, is composed of $n$ pairs of the features' observations and the target's realizations.

    The target and the features are indexed by a multi-index with two levels: (Date $\times$ Stock) as in Example \ref{data_format}. We have structured the data in this way to generate a more robust model and avoid having a specialized model for each stock, which is not efficient.

    \small
    \begin{example}\label{data_format}
    \begin{equation*}
        \begin{matrix}
        Date & Stock & Y \\
            t_1 & S_1 & y_{(t_1, S_1)} \\
            t_2 & S_1 & y_{(t_2, S_1)} \\
            t_3 & S_1 & y_{(t_3, S_1)} \\
            t_1 & S_2 & y_{(t_1, S_2)}  \\
            t_2 & S_2 & y_{(t_2, S_2)} \\
            t_3 & S_2 & y_{(t_3, S_2)} \\
            ... & ... & ...
        \end{matrix}
        \hspace{0.20cm}
        \text{and}
        \hspace{0.20cm}
        \begin{matrix}
            Date & Stock & X[0] & X[1] & ...  \\
            t_1 & S_1 & x^0_{(t_1, S_1)} & x^1_{(t_1, S_1)} & ... \\
            t_2 & S_1 & x^0_{(t_2, S_1)} & x^1_{(t_2, S_1)} & ...  \\
            t_3 & S_1 & x^0_{(t_3, S_1)} & x^1_{(t_3, S_1)} & ...  \\
            t_1 & S_2 & x^0_{(t_1, S_2)} & x^1_{(t_1, S_2)} & ...  \\
            t_2 & S_2 & x^0_{(t_2, S_2)} & x^1_{(t_2, S_2)} & ...  \\
            t_3 & S_2 & x^0_{(t_3, S_2)} & x^1_{(t_3, S_2)} & ...  \\
            ... & ... & ...& ... & ...
        \end{matrix}
    \end{equation*}
    with $y_{(t, s)}$ is the value of the target at time $t$ for the stock $s$ and $x^j_{(t, s)}$ is the observation of the $j$th feature, $X[j]$, at time $t$ for the stock $s$.
    \end{example}
    \normalsize

    \subsection{Rules designing}
        \subsubsection{Rules definition}
        
        A rule is an \emph{If-Then} statement of the form:
        \begin{eqnarray}\label{eq::rule_form}
        	\text{IF} && (X \in c_1) \text{ And }(X\in c_2) \text{ And } \dots \text{ And } (X \in c_k)\\
        	\text{THEN} &&\text{predict the value } p\nonumber
        \end{eqnarray}
        where  $p \in \Rset$ and each $c_i \subseteq \Rset^d$ is expressed in its simplest shape. For instance, the subset $[a_1, b_1] \times [a_2, b_2] \times \Rset^{d-2}$ will be expressed as two $c_i$'s: $$c_1 = [a_1, b_1] \times \Rset^{d-1} \text{ and } c_2 = \Rset \times [a_2, b_2] \times \Rset^{d-2}.$$

        The \emph{If} part, called the \emph{condition} of the rule, or simply the rule, is composed of the conjunction of $k$ basic tests which forms a subset of $\Rset^d$ and $k$ is called the \emph{length} of the rule. The \emph{Then} part, called the \emph{conclusion} of the rule, is the estimated value when the rule is \emph{activated} (i.e., when the condition in the \emph{If} part is satisfied). The rules are easy to understand and allow an interpretable decision process when $k$ is small. For a review of the best-known algorithms for descriptive and predictive rule learning, see \cite{Zhao03} and \cite{Furnkranz15}. 

        It is important to notice that a rule is completely defined by the hyper-rectangle formed by its condition. Thus, by an abuse of notation, we do not distinguish between a rule and the hyper-rectangle.

        \subsubsection{RuleSet definition}
        We denote $R_{\{s\}}$ the set of all rules implying features in the sequence $s$. For example, $R_{\{i, j\}}$ refers to the rules with conditions on the features $i$ and $j$. 
        
        \subsubsection{Activation vector}
        
        We call the \textit{activation vector}, the vector of length $n$ formed by the evaluation of the condition on the observations of the features on $D_n$ :
        it contains 1 at each point in the index when the condition is met by its subset of $X$ and 0 elsewhere.

        \subsubsection{Discretization process}
        

        Each continuous variable of the feature space is discretized into $m_n$ \textit{bins}. As described in \cite{margot2021esg} the empirical quantiles of each variable are considered. The algorithmic complexity of the discretization is $O(ndm_n)$. 
        
    	Using discretization is a common method when learning on continuous variables, we may cite 
    	the algorithms \emph{BRL} (Bayesian Rule Lists) \cite{Letham15}, RIPE \cite{Margot18} and \emph{SIRUS} \cite{Benard21} which also use the empirical quantiles.

        \begin{remark}
        One can notice that features may have different distribution given the stock. In this case the discretization process must be done considering each stock distribution separately. One way to get around this is to generate features through a data engineering process to get features without units. In this paper, we have chosen this last option.
        \end{remark}

        \subsubsection{Designing rules of length 1}
        
        We distinguish two meaningful ways of generating the rules : \textit{full-interval} and \textit{half-interval}, the latter creating fewer rules, rules that, in some cases, will be more interpretable.

        \paragraph{\textbf{Full-interval}:} This consists in considering a condition with consecutive bins for a given feature (i.e. one given feature can appear at most once in a given condition), except the trivial one $c_1=[0, m_n-1]$\footnote{We index the bins from $0$ to $m_n-1$}.
        Thanks to the discretization we have at most $\dfrac{(m_n+1)m_n}2-1$ rules by feature, so the number of rules of length $1$ is
    	\begin{equation}\label{eq::nrules_1_FI}
    	    \R_{1_{FI}} = d\times \sum_{k=1}^{2} \binom{m_n}{k} - 1=d\times \left(\frac{m_n(m_n+1)}{2}-1\right).
    	\end{equation}
    	From this it is easy to see that the complexity of evaluating all rules of length $1$ using full-interval is $O(ndm_n^2)$. The linear dependence on $n$ comes from the fact that a rule is defined by its activation vector of length $n$, and that evaluating this activation vector is linear in $n$.
    	
    	\paragraph{\textbf{Half-interval}:} This consist in considering conditions with consecutive bins AND with one bin being $0$ or $m_n-1$. This approach generates a set of rules easier to translate into natural language. 
    	Using half-interval the number of rules of length $1$ becomes,
    	\begin{equation}\label{eq::nrules_1_HI}
    	    \R_{1_{HI}} = 2\times d \times (m_n - 1).
    	\end{equation}
    	It is easy to see that the complexity of evaluating all rules of length $1$ using half-interval is $O(ndm_n)$. So in addition to being more interpretable, half-intervals are also much quicker than full-intervals.

        \subsubsection{Designing rules of length $2$}
        
        A rule of length $2$ can be seen as one rule of length $1$ implying the feature $X[i]$ and one rule of length $1$ implying a different feature. Generating all rules of length $2$ consists of considering all the pairs of rules of length $1$ that imply different features. To generate a new rule from two other rules, it is enough to compute a \emph{logical and} of the two activation vectors of the two rules to create the activation vector of the new rule.

        
         We have $\# R_{\{i\}} := \dfrac{\R_1}{d}$ rules of length $1$ implying the feature $X[i]$. And we have $\frac{\R_1}{d} \times (d-1)$ rules of length $1$ implying features different from $X[i]$. Hence, the number of rules of length $2$ implying two different features is $\left(\frac{\R_1}{d}\right)^2\times (d-1)$. So, the number of rules of length $2$ is
        
        \begin{equation}\label{eq::nrules_2}
            \R_2 = \left(\frac{\R_1}{d}\right)^2\times \sum_{i=1}^{d} \left(d-i\right)=\left(\frac{\R_1}{d}\right)^2 \frac{d(d-1)}{2}.
        \end{equation}

        We have $\# R_{\{i, j\}} := (\R_1/d)^2$ rules implying the same pair of features and $d(d-1)/2$ rulesets. From this we can see that the complexity of evaluating all rules of length $2$ using half-interval is $O(n\R_1^2)$.

        \subsubsection{Designing rules of length $l>2$}\label{sec::nrules_l}
        
        Each rule of length $l-1$ can be combined with rules of length $1$ implying features not implying in the condition of the first rule. 
        From one rule of length $l-1$, there are $\frac{\#R_1}{d}\times (d-l+1)$ new rules of length $l$. So the number of rules of length $l$ is
    
        \begin{equation}\label{eq::nrules_l}
            \R_l = \frac{\R_{l-1}\times \R_1}{d\times l}\times (d-l+1).
        \end{equation}
        
        The $l$ in the denominator in \eqref{eq::nrules_l}, comes from the redundancies when generating rules of length $l$. 
        From this, we can notice that the complexity of evaluating all rules of length $l$ using half-interval is $O(n\#R_{l-1} \#R_1)$.
        
        Finally, the total number of rules is bounded by:
        \begin{equation}\label{eq::nrules}
            \R = \sum_{l=1}^{l_{max}}\R_l = \sum_{l=1}^{l_{max}}\# R_{\{s(l)\}} \times \#\,\left\{R_{\{s\}},\, s \in \P_l\left(\{1, \dots, d\}\right) \right\},
        \end{equation}

        where $s(l)$ is a sequence of $l$ elements from $\{1,2, \dots, d\}$ and $\P_l(\{1, \dots, d\})$ is the set of all subsets of $l$ elements from $\{1, \dots, d\}$.
        
        The terms $\# R_{s(l)}$ is the number of elements in any ruleset of rules of length $l$ and $\#\,\left\{R_{\{s\}},\, s \in \P_l\left(\{1, \dots, d\}\right) \right\}$ is the number of ruleset of rules of length $l$. We have

        \begin{align}
             \#\,R_{s(l)} & =\left(\frac{\#R_1}{d}\right)^l, \\
             \#\,\{R_{\{s\}},\, s \in \P_l\left(\{1, \dots, d\}\right) \} & =\prod_{j=1}^l \left(\frac{d-(j-1)}{j}\right).
        \end{align}
        The algorithm complexity is then $O(n(dm_n)^l)$ in the setting of half-intervals, and  $O(n(dm_n^2)^l)$ in the setting of full-intervals.

        \subsubsection{Important attributes}
        
        In addition to the rules designing, the algorithm also calculates several attributes to sort and filter them. Computing those attributes is called \textbf{fitting} the rule. It is the most time-consuming part of a rule designing process. We have four main attributes:
            \begin{itemize}
                \item The prediction, which is the conclusion, is the equal to the mean of all $Y$ on activated points.
                \item The coverage rate, or simply the coverage, denoted $c(r_k, D_n)$, defined as the ratio between the number of activation points and the length $n$ of the index.
                \item The criterion, which is chosen according to the settings of the problem, denoted $crit(r, D_n)$. It is computed in a rolling window and returns a sequence. 
                \item The significant test, which evaluates the statistical significance of the rule with a statistical test based on a concentration measure.
            \end{itemize}
    	
        In AdLearn, a rule must fulfil four conditions :
        \begin{enumerate}
            \item $c(r_k, D_n) > cov\_min$.
            \item $Average(crit(r_k, D_n)) > K$ if we want to maximize the criterion or $crit(r_k, D_n) < K$ if we want to minimize it. 
            \item $|\min\left(crit(r_k, D_n)\right)| < |\max\left(crit(r_k, D_n)\right)|$.
            \item The significant test must be fulfilled with a p-value greater than $0.05$.
        \end{enumerate}
        
        
        \subsubsection{Selection process}
        
        Unlike the original RIPE algorithm \cite{Margot18} or its first update for financial forecasts presented in \cite{margot2021esg}, this selection process is not based on optimizing the criterion. To limit the computation time, we have adapted the selection process of CoveringAlgorithm presented in \cite{margot2021consistent}. The rules are browsed according to the sorting criterion, from the best to the worst. The best rule is added to the selected set of rules (initially empty) $S_n$ and the next rule is added if and only if it has at least a rate $1 - \gamma$, for a chosen $\gamma \in (0, 1)$, of points not covered by the current set of selected rules. The selection process stops when all rules have been browsed one time. The process is described in pseudocode in Algorithm \ref{algo::selection}.

        \begin{algorithm}
        	\caption{Selection of minimal set of rules}
        	\label{algo::selection}
        	\SetAlgoLined
        	\KwIn
        	{
        		\begin{itemize}
        			\item the rate $0<\gamma<1$;
        			\item a set of rules $RS_n$;
        		\end{itemize}
        	}
        	\KwOut
        	{
        		\begin{itemize}\vspace{-0.2cm}
        			\item a selected set of rules $S_n$;
        		\end{itemize}
        	}
        	$RS_n \leftarrow sort\_by\_crit(RS_n)$\;
        	$S_n \leftarrow \{\emptyset\}$ \;
        	\For{$r'$ in $RS_n$}
        	{
        		\uIf{ $c(r' \cap \{\cup_{r \in S_n} r\}, D_n) \le \gamma\,c(r', D_n)$}
        		{
        			$S_n \leftarrow S_n \cup r'$\;
        		}
        	}
        
        	\Return{} $S_n$\;	
        \end{algorithm}
        \normalsize

    \subsection{Input used in this paper}\label{sec::inputs}
    \begin{itemize}
        \item Number of bins $m_n$: We chose $5$ bins in order to be able to interpret our predictions in terms of \textit{very small}, \textit{small}, \textit{average}, \textit{large}, and \textit{very large} values of features.
        \item Index size $n$: We used an index that has $1974672$ points : more than $8$ years and more than $1000$ stocks. 
        \item Number of features $d$: $1238$, being about financial and extra-financial notes.
        \item Criterion: In this application, we have chosen a criterion that is specifically suited for financial applications. 
        \item Thresholds: $cov\_min = 0.05$ and $K = 0$.
        
        \item Intervals: Full.
        \item The maximum length of the rules $l_{max}$: 2.
    \end{itemize}
    
\section{Implementation}

    \begin{figure}[ht!]
        \centering
        \includegraphics[width=\textwidth]{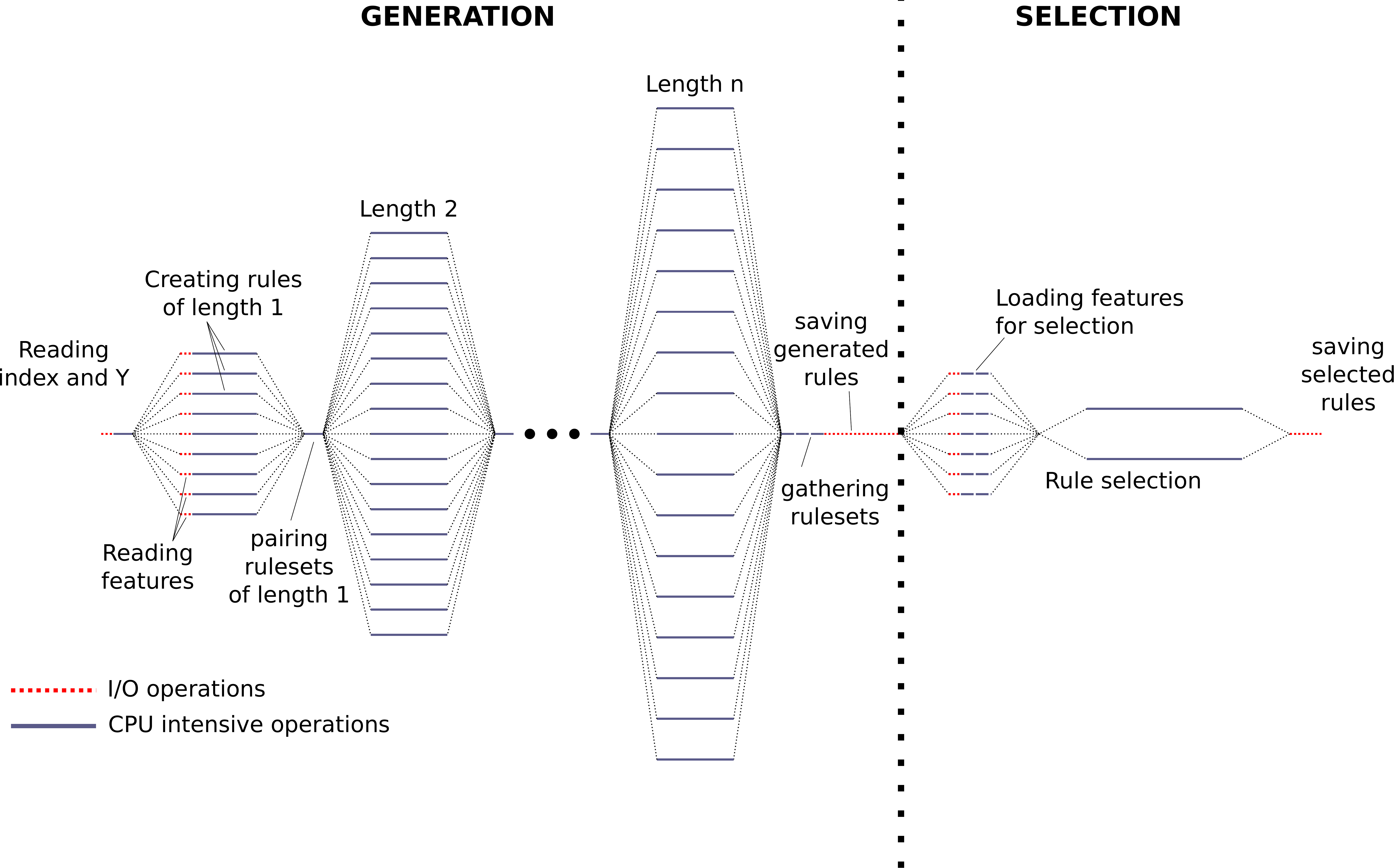}
        \caption{Representation of AdLearn algorithm. It is separated in two parts that can be run separately: the generation and the selection. The generation is separated in steps. Each step corresponds to the generation of all the rules of a given length. One step can be parallelized: rulesets of a given length are generated in different processes (see Fig.~\ref{fig::parallel}), so multiprocessing will be efficient as long as the number of rulesets is greater than or equal to the number of cores. Each step needs the results from the previous step to operate, thus the presence of bottlenecks between steps. However, in a single step, the processes are independent and do not communicate with each other.}
        \label{fig::adleran} 
    \end{figure}

    The algorithm is implemented in Python and can run on Python $\ge3.7$ and Intel Distribution for Python 3.7.11. It uses the \textit{Pandas} library for the computations, the \textit{multiprocessing} library for the parallelization, the \textit{Ruleskit} library for handling the conditions, rules, activation vectors and rulesets objects. And the \textit{transparentpath} library to read and to write data on Google Cloud Storage. Because \textit{Pandas} uses locks, the parallelization context used is \textit{spawn} instead of the default \textit{fork}. Fig.~\ref{fig::adleran} shows a representation of the algorithm.
    
    \subsection{Memory usage}
    
        Among all objects loaded or created by the algorithm, the most memory consuming are $Y$, $X$ and the activation vectors. $Y$ contains $n$ floating-point numbers, $X$ contains $d\times n$ floating points numbers and an activation vector contains $n$ boolean values.

        $Y$ and $X$ are not bottlenecks: $Y$ ($\sim \SI{16}{\mega\byte}$ in our case) is needed once per process. Only one $X[i]$ (same size as $Y$) per process is needed, and only to generate rules of length $1$. The activation vectors however could be problematic if we keep them all in memory when generating a lot of rules. In order to both saving memory and reducing overhead when scheduling processes, they are written to disk instead and read when needed, passing only string objects to the various processes generating the rules.
        
        To allow for efficient parallelization, one must pay attention to the amount of memory a single process is expected to use. In our case, with the criterion function we use, each process needs $\sim 40\times n\sim\SI{80}{\mega\byte}$.
        
    \subsection{Reduction of overhead}
        To reduce overhead, the methods creating the rules take as argument small objects, the bigger being a list of length $d$ (1238 in our case). $Y$ and the activation vectors are written on disk in compressed format, and only paths to those are passed to the processes. The processes do not need to communicate with each other.
        
        We found that if computing one rule per process, even with the effort made to reduce the overhead due to scheduling the processes and getting back their results, the run time was greater than when using one process to compute one ruleset.
        
        We also found that the load unbalance was great if we asked one process to compute $\#\,\text{cores}/\#\,\text{pairs}$ rulesets instead of one. Eventually, we made one process produce one ruleset.

    \subsection{Rules generation}
    
        From equations \eqref{eq::nrules_1_FI} and \eqref{eq::nrules_2} and section \ref{sec::inputs}, we know that we expect to generate $$\R_1 = \num{14} \text{ rules} \times \num{1238} \text{ rulesets} = \num{17332} \text{ rules}$$ $$\R_2 = \num{196} \text{ rules} \times \num{765703} \text{ rulesets} = \num{150077788} \text{ rules}$$ for a total of $\R=\num{150095120}$ rules.
        
        Knowing that generating a rule takes around 2\,s, computing all those rules would take a bit less than 10 years. However, the great number of individual items allows for a massive parallelization of the code.
        
        \begin{wrapfigure}{r}{0.5\textwidth}
            \centering
            \includegraphics[width=0.5\textwidth]{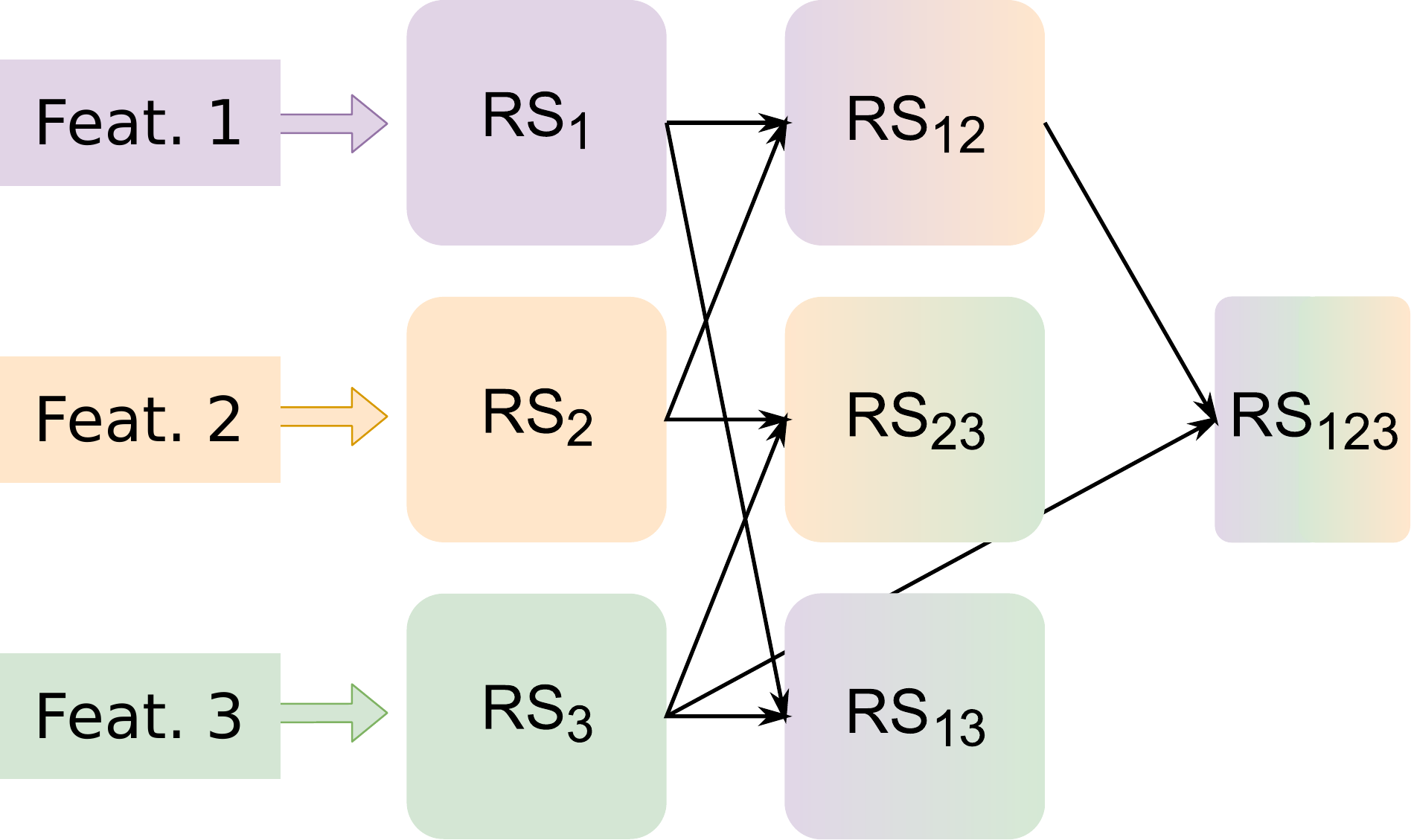}
            \caption{How the parallelization of the code is achieved. Each ruleset (RS) is generated in a process. Example with 3 features.}
            \label{fig::parallel}
        \end{wrapfigure}
        
        In practice, the code is parallelized on the rulesets, as shown in Fig.~\ref{fig::parallel}, whereas the version of \cite{margot2021esg} is parallelized on the designing of the rules only\footnote{In practice, this requires too much RAM. In fact, each rule must have its activation vector and $Y$ in memory, whereas the new version requires one $Y$ per ruleset.}. One can note from Fig.~\ref{fig::parallel} that, as mentioned in section \ref{sec::nrules_l}, the ruleset of length $3$, $R_{\{1,2,3\}}$ could be generated using $R_{\{1\}}$ and $R_{\{2, 3\}}$, $R_{\{2\}}$ and $R_{\{1, 3\}}$ or $R_{\{3\}}$ and $R_{\{1, 2\}}$. In order to do the computation only once, each feature is associated to an index, and only processes combining $R_{\{i\}}$ with $R_{\{j, k\}}$, with $j\ne i\ne k$ and $i>\max(j,k)$ can be scheduled.
        
        In addition to the parallelization, the fact that thresholds are used will decrease the number of generated rules, also decreasing the computation time. For a given rule, each time an attribute is computed, the program checks whether a threshold exists for this attribute. If so and if the rule does not meet it, the generation stops for this rule. Furthermore, we assume that a "bad" rule of length 1 will not generate a "good" rule of length 2 or more. We can see from equation \eqref{eq::nrules_2} that if only 10\,\% of $\R_1$ are left, at most 1\,\% of $\R_2$ can be generated. In our practical case, \num{230} rules of length $1$ and \num{5572} rules of length 2 are left, for a total of \num{5802} rules. \num{8911} rulesets of length 2 were generated (some ended up empty), so we should be able to make use of a machine with up to $\sim 10$\,k cores. We will see in section \ref{sec::extrapolation} that this is however not the case. Note that we do not create pairs of two rulesets if at least one is empty.
        
        \begin{remark}
            It can be noted here that the method described above to avoid computing several times the same ruleset will produce exactly the same results no matter how we choose to compute this ruleset when no thresholds are used. If using thresholds however, results can vary. Indeed, in the example of Fig.~\ref{fig::parallel}, if $R_{\{1, 2\}}$ ends up having no good rules, then $R_{\{1, 2, 3\}}$ will be empty, while it might not be had we chosen to generate it with $R_{\{1\}}$ and $R_{\{2, 3\}}$ instead.
        \end{remark}

        The output of the rules' generation is a CSV file containing one rule per line, and one column per rule attribute. A sample of such a file is shown in Table~\ref{tab::output}
        
        \begin{table}[htp!]
        \centering
            \caption{Example of CSV file created by the rules generation.}
            \label{tab::output}
            \begin{tabular}{l|lllllll}
                     & names & bmins & bmaxs & cov & pred & ... \\ \hline
            r\_0(2)+ & ['X[1]'] & [0] & [1] & 0.39 & 0.012 & ... \\
            r\_1(2)- & ['X[1]', 'X[2]'] & [3, 0] & [4, 1] & 0.11 & -0.054 & ... \\
            r\_2(2)+ & ['X[2]', 'X[3]'] & [0, 2] & [1, 3] & 0.12 & 0.051 & ... \\
            r\_3(2)- & ['X[2]', 'X[3]'] & [0, 2] & [1, 3] & 0.15 & -0.049 & ... \\
            r\_4(2)- & ['X[2]', 'X[4]'] & [1, 2] & [2, 4] & 0.16 & -0.049 & ... \\
            ... & ... & ... & ... & ... & ... & ...        
            \end{tabular}
        \end{table}
    
    \subsection{Rules Selection}
    
        The selection can be done separately from the generation. It will need to recompute the activation vectors of each generated rules, for they are automatically deleted from the disk after the generation. That is not very time-consuming since it can be parallelized, and that the number of rules is limited when using thresholds (\num{5802} in our case).
        
        The selection process itself, described in algorithm \ref{algo::selection}, is done separately on rules with a positive prediction and those with a negative prediction. This is done for financial purposes, since those two kinds of rules will create respectively buying or selling recommendations. If the algorithm is used in another context, it might not make sense to do it that way. The two resulting sets of rules are then concatenated, and the output will be the same kind as the one shown in Table~\ref{tab::output}.
        
        In its present form, the selection algorithm \ref{algo::selection} can not be parallelized. Indeed, the algorithm needs the result of the previous step to take the next step.
        
        In our practical case, the generation kept 21 rules (the version presented in \cite{margot2021esg} kept 18 rules). That is small enough to be interpretable.


\section{Computation times}


    \subsection{Comparing various machines}
        
        We studied the generation time of the algorithm on different architectures :
        \begin{itemize}
            \item GCP E2 Standard VM with \SI{16}{vCPUs} Intel(R) Xeon(R) CPU @ \SI{2.2}{\giga\hertz} and \SI{128}{\giga\byte} memory.
            \item GCP N2D Custom VM with \SI{64}{vCPUs} AMD EPYC 7B12 @ \SI{2.25}{\giga\hertz} and \SI{512}{\giga\byte} memory.
            \item GCP N2D High-CPU VM with \SI{224}{vCPUs} AMD EPYC 7B12 @ \SI{2.25}{\giga\hertz} and \SI{224}{\giga\byte} memory.
            \item \textit{Bisocket} server provided by Fujitsu Europe with \SI{2}{CPUs} Intel(R) Xeon(R) Gold 6240 @ \SI{2.6}{\giga\hertz} (72 threads in total), \SI{192}{\giga\byte} memory and \SI{1.8}{\tera\byte} SSD.
        \end{itemize}
        
         The generation times are shown in Table~\ref{tab::time_perf}, where we limited the multiprocessing on the \textit{bisocket} machine to 64 processes for comparison purposes with the \textit{64 vCPUs} GCP VM, since VMs on GCP can only have a number of cores that is a multiple of 16. All GCP VMs have \SI{1}{\tera\byte} SSD. Those configurations allow for storing up to one million activation vectors on disk, and have more than enough memory per core. From this table we can see that, if using the same number of cores, the bisocket server is 23\,\% quicker than the GCP VM when using standard Python 3.8, and 28\,\% quicker when using Intel Python. Using Intel Python with 64 cores on the bisocket machine provides an acceleration of  15\,\%.
        
        \begin{table}[ht!]
            \centering
            \begin{tabular}{|l|cccc|cccc|}
                \hline
                 & \multicolumn{4}{c|}{Python 3.8} & \multicolumn{4}{c|}{Intel Python} \\
                \cline{2-9}
                & \multicolumn{1}{c}{\textbf{Bisocket}} & \multicolumn{1}{c}{\textbf{\begin{tabular}[c]{@{}c@{}}GCP\\ 16\end{tabular}}} & \multicolumn{1}{c}{\textbf{\begin{tabular}[c]{@{}c@{}}GCP\\ 64\end{tabular}}} & \multicolumn{1}{c|}{\textbf{\begin{tabular}[c]{@{}c@{}}GCP\\ 224\end{tabular}}} & \multicolumn{1}{c}{\textbf{Bisocket}} & \multicolumn{1}{c}{\textbf{\begin{tabular}[c]{@{}c@{}}GCP\\ 16\end{tabular}}} & \multicolumn{1}{c}{\textbf{\begin{tabular}[c]{@{}c@{}}GCP\\ 64\end{tabular}}} & \multicolumn{1}{c|}{\textbf{\begin{tabular}[c]{@{}c@{}}GCP\\ 224\end{tabular}}}\\
                \hline
                \textbf{Time (s)} & \num{3369} & \num{16638} & \num{4397} & \textcolor{jd_green}{\num{1691}} & \num{2849} & \num{11494} & \num{3962} & \textcolor{jd_green}{\num{1309}} \\
                \textbf{\begin{tabular}[c]{@{}l@{}}Time per\\ rule (s/rule)\end{tabular}} & \num{0.58} & \num{2.9} & \num{0.76} & \num{0.29} & \num{0.49} & \num{1.98} & \num{0.75} & \num{0.23} \\
                \textbf{\begin{tabular}[c]{@{}l@{}}Time $\times$ \#\,CPU\\ ($10^5$ \,s$\cdot$CPU)\end{tabular}} & \textcolor{jd_green}{\num{2.2}} & \num{2.7} & \num{2.8} & \num{3.8} & \textcolor{jd_green}{\num{1.8}} & \textcolor{jd_green}{\num{1.8}} & \num{2.5} & \num{2.9} \\ 
                \hline
            \end{tabular}
            \caption{\label{tab::time_perf}Run times of AdLearn (inputs shown in section \ref{sec::inputs}) on various machines, using Python 3.8 (top) and Intel Python 3.7 (bottom). For comparison, the version presented in \cite{margot2021esg} took 14\,h for the same run on a 16 cores machine. The run time has thus been divided by 4.4 on the same architecture (code optimization only) and by 39 when using 224 cores.}
        \end{table}

    \subsection{Quality of the learning}
        We compare here the quality of the learning obtained with the new version of AdLearn, which does not use candidates to reduce the number of rules, to the quality of the learning obtained with the algorithm described in \cite{margot2021esg}, which uses only 40 rules of length 1 to generate rules of length 2.
        
        The version with candidates was designed to run in a reasonable time on a desktop computer, at the price of exploring less rules. The new version aims at using HPC to explore more rules in a reasonable time. We created two simple investment strategies.
        
\begin{wrapfigure}[18]{r}{0.5\textwidth}
            \centering
            \includegraphics[width=0.5\textwidth]{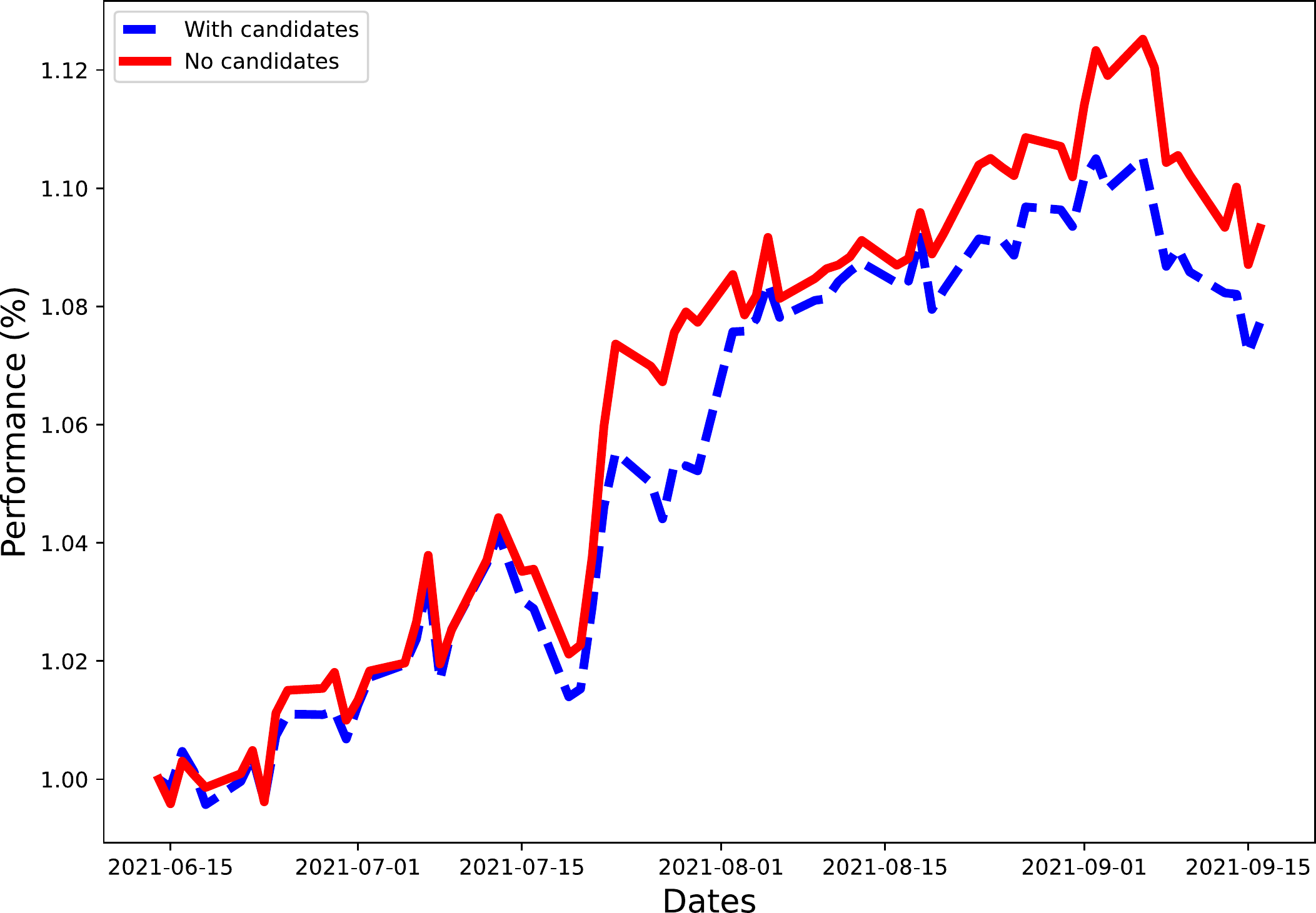}
            \caption{Compared cumulated predictive performances of the new version of AdLearn (labelled 'No candidates') vs the version presented in \cite{margot2021esg} (labelled 'with candidates'). On this particular time frame, the new version shows a consistently higher predictive rate.}
            \label{fig::perf}
        \end{wrapfigure}

        For each generated ruleset, the predictions of the rules are aggregated, as in \cite{margot2021esg}, to obtain a vector of predictions for each ruleset. On the first business day of each month, the 40 stocks with the highest predictions in each prediction vector are selected. These stocks are equally weighted and form two portfolios, one for each ruleset. The composition of the portfolios may change each month. Finally, to evaluate the performance, we calculate the cumulative average daily return of the stocks in the portfolios. These results are shown in Fig.~\ref{fig::perf}.

        We can see a net enhancement of the performance when exploring more rules. This result was obtained in less than one hour using Fujitsu's bisocket server with 72 cores and the thresholds presented in section \ref{sec::inputs}, and comforts us in the idea that using more permissive thresholds (see section \ref{sec::extrapolation}), especially on the coverage, would allow us to enhance the learning even more and would benefit from a larger number of cores.
        
    \subsection{Extrapolation to very high number of cores}\label{sec::extrapolation}
    
    For this paper, we were limited to 72 cores on the bisocket machine. Since Fujitsu can provide much bigger machines (several thousands of cores), it is interesting to extrapolate the run time to very high number of cores, to see if there can be a significant gain in using thousands instead of 72 cores.
    
    We assume that generation time $t_G$ is a function of the number of cores that follows Amdahl's law\cite{amdahl}:
    \begin{equation}\label{eq::amdahl}
        t_{G}(\#\,\text{cores}) = \left((1-p) + \frac{p}{\#\,\text{cores}}\right)\times t_{G}(1),
    \end{equation}
    where $t_{G}(1)$ the generation time in a serial run
    and $p$ the fraction of the code that benefits from the multiprocessing.
    The generation time is directly measured and $p$ is a parameter that is adjusted to the measurements. The use of Amdahl's law instead of Gustafson's law\cite{Gustafson88} is justified by the fact that Gustafson's law supposes a fixed run time, and an optimisation of the algorithm to use all the available cores at their maximum, while in our case the algorithm does not change when the number of cores changes, matching the condition to use Amdahl's law.
    
    
    
\begin{figure}
        \begin{minipage}{0.6\textwidth}
        \includegraphics[width=\textwidth]{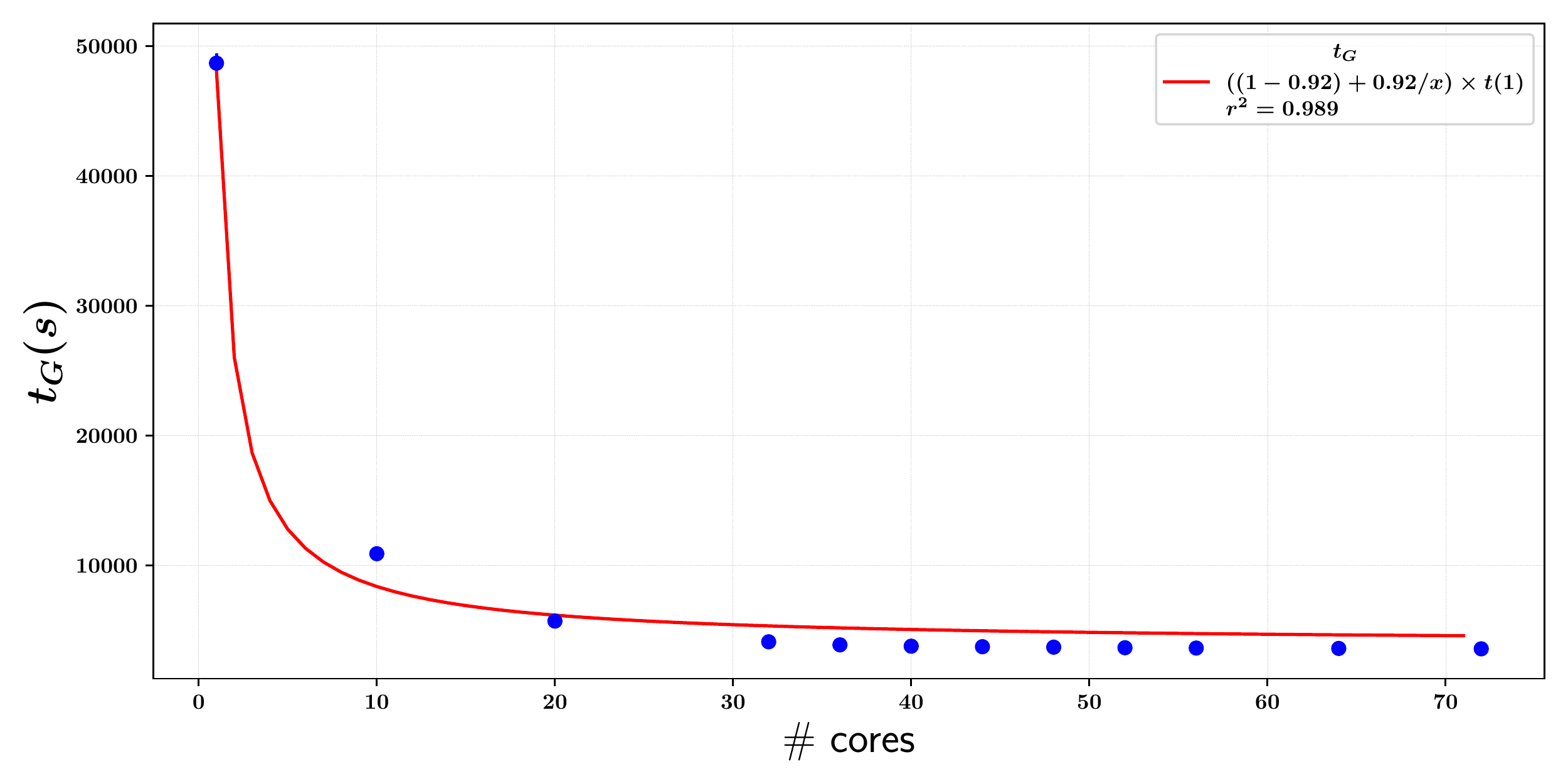}
        \caption{Generation time $t_G$ against the number of cores and the fit of equation \eqref{eq::amdahl} to it. One can note that the time taken with one core ($\mathbf{\sim}$14\,h) is equivalent to the time taken by the version presented in \cite{margot2021esg} (see Table~\ref{tab::time_perf}), which is parallel too, meaning that the algorithm was efficiently optimised.}
        \label{fig::time_fit}
        \end{minipage}\hfill
        \begin{minipage}{0.38\textwidth}
            \includegraphics[width=\textwidth]{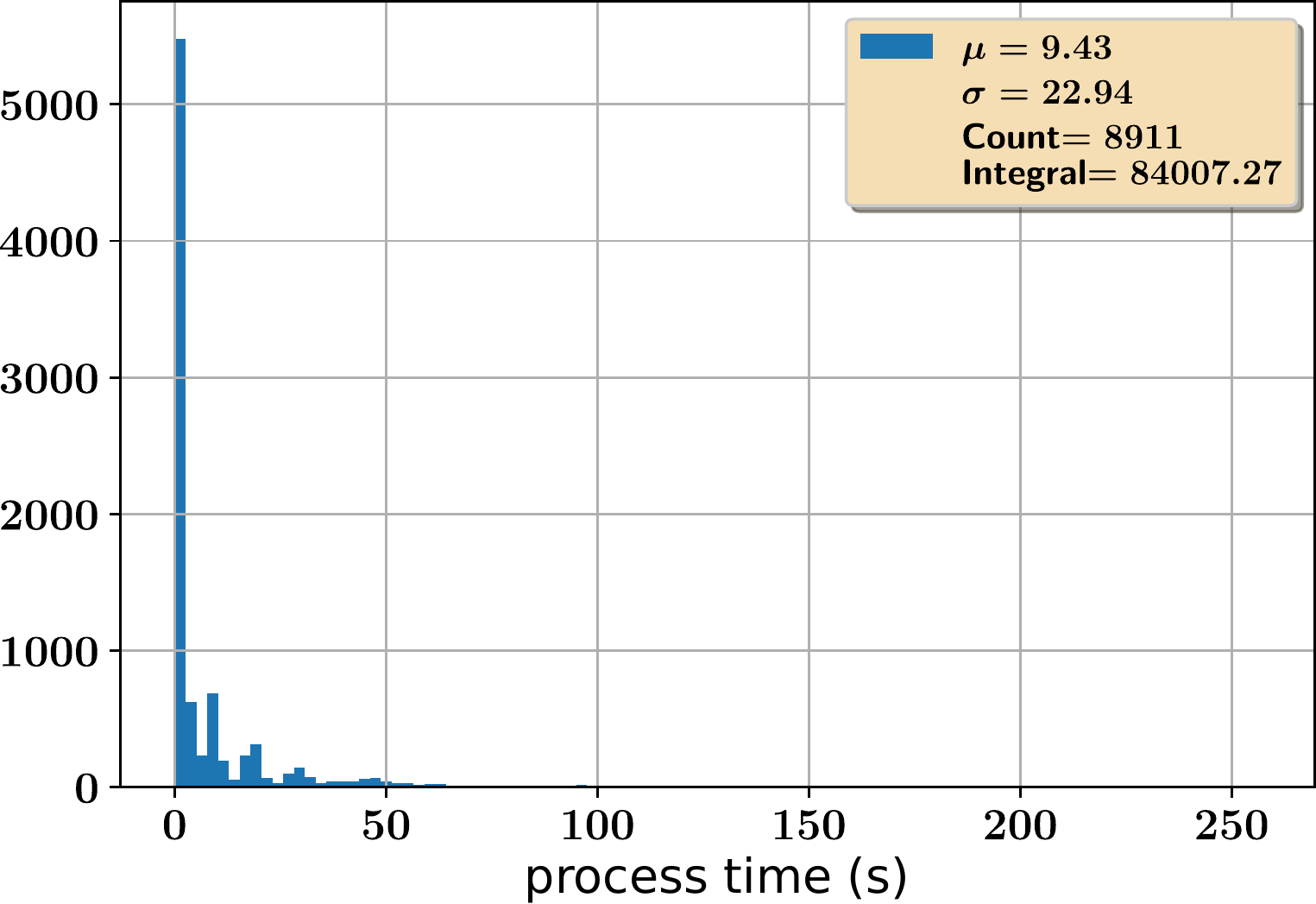}
            \caption{Distribution of the run times of individual processes generating rules of length 2. One can see an accumulation of very fast processes, corresponding to rulesets with very few "good" rules, and a bump of processes taking $\sim$\,50\,s.}
            \label{fig::times_histogram}
        \end{minipage}
    \end{figure}

    Fig.~\ref{fig::time_fit} shows $t_G$ against the number of cores used, and the fit of equation \eqref{eq::amdahl} to it. We can see that according to Amdahl's law, 92\,\% of the rules generation benefits from parallel processing. Four runs were done for each point to provide statistics, but the variations of the generation time are too small to be visible on the graph.
    
    One can clearly see that after 30--40 cores, the gain in generation time is very limited. This can be explained by Fig.~\ref{fig::times_histogram}, which represents the distribution of the run time of each individual processes generating the rules of length 2. One can see that most of them finish very quickly, while only a few take a significant time ($\ge \SI{50}{\second}$). This is because we use thresholds to filter out "bad" rules : those thresholds will remove most of the rules (only 5570 rules of length 2 are left when more than $150\times 10^6$ are expected), making a lot of rulesets empty or almost empty. Those rulesets will compute very fast, for a rule does not need to complete its entire fit to be detected as bad (the threshold on coverage for example only needs the activation vector to be evaluated). And only around thirty to forty rulesets contain a significant amount of good rules.
    
    \subsection{Future work : time against number of features, index length and number of cores}  
    
    The results presented in the previous section are valid for this particular study, but another run, using less harsh thresholds, could find completely different results. Thus, it is still interesting to investigate the generation time against the number of cores when using no thresholds. This can hardly be done using all features however, since that would imply generating all the $150\times 10^6$ rules, which would take too long, even more so if we want to iterate on the number of cores while doing that. Instead, we can study the generation time using a smaller index and iterating on the number of cores and the number of features, starting at 16 (fewer features imply too few rulesets to efficiently use many cores), and the computation time can then be extrapolated to 1238 features.
    
    This study will be presented in a future work.

\section{Conclusion and future works}
We have designed an interpretable predictive algorithm and executed it on an HPC machine. It appears that the limiting factor for taking advantage of a large number of cores is the importance of thresholds used to limit the exploration of the learning. The effect of these thresholds can be reduced at will, in order to explore more rules. We have also found that widening the space of explored rules enhances the performance of the investment strategy created from the learned ruleset. This is in line with an intuitive expectation linking a better financial performance to a higher variety of economical facts retained in the prediction.

The next step will be to study in details the behavior of the algorithm against the number of cores when no thresholds are used, to estimate the number of rules we could generate in a reasonable amount of time given a number of cores.

Also, this paper did not look into the selection step, that follows the generation. Although the numbers are not presented in this work, we observed that in its current implementation, the time spent in the selection process is very sensitive to inputs, being in general of the same order as the generation time. This clearly constitutes a potential bottleneck, given the polynomial nature of the current selection algorithm.
A redesign of this algorithm appears necessary, with comparable hopes for improvement as for the exploration part. Simulated annealing techniques, and maybe even quantum annealing, are very likely to be good candidates to tackle this issue.

\bibliography{ref}
\bibliographystyle{splncs04}

\end{document}